\documentclass[prl,twocolumn,showpacs,graphicx]{revtex4}
\usepackage{amsfonts}
\usepackage{amssymb}
\usepackage{amsmath}
\usepackage{graphicx}

\begin{document}

\title{Semiconducting graphene nanomeshes}
\author{I.I. Naumov and A.M. Bratkovsky}
\affiliation{Hewlett-Packard Laboratories, 1501 Page Mill Road, Palo Alto, California
94304}
\date{\today}
\keywords{}

\begin{abstract}
Symmetry arguments are used to describe all possible two-dimensional
periodic corrugations of graphene (``nanomeshes") capable of inducing
tangible semiconducting gap. Such nanomeshes or superlattices break the
initial graphene translational symmetry in a way that produces mixing and
subsequent splitting of the Dirac $\boldsymbol{K}$ and $\boldsymbol{K}%
^{\prime }$ states. All of them have hexagonal Bravais lattice and are
described by space groups that are subgroups of the graphene group. The
first-principles calculations show that the gaps of about 0.5~eV can be
induced at strains safely smaller than the graphene failure strain.
\end{abstract}

\pacs{73.22.Pr,  81.05.ue, 62.25.-g}
\maketitle

Graphene, a truly two-dimensional crystal of carbon atoms with a honeycomb
lattice, exhibits giant carrier mobilities up to $6 \times 10^6$ cm$^2$/Vs
\cite{bolotin,morozov,dusari2011}, it has a record mechanical strength with
critical deformations exceeding 20\% and huge thermal conductivity reaching
1800 WK$^{-1}$m$^{-1}$\cite{leeThrmlG}. As such, it is very promising as a
circuit material in electronic devices. However, the absence of a band gap
in graphene prevents its immediate use as e.g. a channel in switchable
devices like field effect transistors.

Perhaps, the most straightforward way to induce a gap in graphene is through
quantum confinement - one should simply cut a graphene sheet into small
pieces like nanoribbons \cite{berger,han} or dots \cite{ponomarenko}.
However, it is difficult to control the band gap produced in such a way
because it strongly depends on the size and edge geometry of the
nanostructures. That is why many papers have been focused on the opening up
a spectral gap in a `bulk' material using different kinds of \emph{periodic}
lattice perturbations with various substrates \cite{zhou,gioavannetti},
antidot lattices \cite{yu,vanevic,liu,baskin}, regular lines or areas of
adsorbed foreign atoms \cite{chernozatonskii,ito,balog,xu}, external
periodic electric and magnetic fields \cite{snyman}, laser field \cite%
{savel'ev}, Kekul\'{e} distortions \cite{lee}, etc.

Mechanical out-of-plane sine-wave deformations certainly belong to this
class of perturbations. Recently, we showed that the one-dimensional (1D)
periodic ripples can open a gap at the Dirac points only due to (i) breaking
of the inversion symmetry or equivalency between A and B sublattices and/or
(ii) merging of two inequivalent Dirac points, $\boldsymbol{D}$ and $-%
\boldsymbol{D}$ \cite{naubra1d}. Breaking of inversion symmetry was found to
have only a relatively modest contribution to the gap, especially when the
amplitude $A$ of a sine wave deformation is relatively small in comparison
with the period $\lambda $. A tangible gap can mainly result from mutual
annihilation of the Dirac points $\boldsymbol{D}$ and $-\boldsymbol{D} $,
yet this requires severe corrugations.

Importantly, in the case of 1D periodic corrugations the energy gap at the
Fermi level cannot be induced due to mixing of electronic states belonging
to two different valleys (or to two inequivalent Dirac points $\boldsymbol{K}
$ and $\boldsymbol{K}^{\prime }$ in the unperturbed graphene) even if the
imposed period provides a momentum transfer $\boldsymbol{g}=\boldsymbol{K}-%
\boldsymbol{K^{\prime }}$ coupling the Dirac points. The reason is that the
Dirac points generally shift from their initial positions as the amplitude
of corrugations $A$ increases, so that the relation $\boldsymbol{g}=%
\boldsymbol{K}-\boldsymbol{K^{\prime }}$ does not hold any longer at any
finite amplitude $A$.

In the present study, we show that in contrast to 1D the 2D out of plane
modulations can easily induce the gap at the Dirac points via $\boldsymbol{K}%
-\boldsymbol{K^{\prime }}$ mixing, provided that the modulations have
hexagonal translational symmetry. It might appear that this idea falls under
the concept developed in Ref.~\cite{pacoNP10} of creating the gaps in
graphene by putting the latter on top of a corrugated surface with a
triangular landscape. In reality, however, there is nothing in common
between the two since we consider the \emph{semiconducting} gaps forming at
the Dirac points due to inter-valley mixing, while the authors \cite%
{pacoNP10} studied the multiple minigaps lying above (and below) the Dirac
energies.

The opportunity to open up the gap through 2D corrugations has become
especially interesting in light of the recent experiments showing that such
corrugations ("nanomeshes") can be formed during the epitaxial graphene
growth on some special substrates like SiC(0001) \cite{emtsev,kim1,riedl,qi}
or transition metal surfaces \cite{preobrajenski,martoccia,ma}. When grown,
for example, on Ru(0001), graphene forms a nanomesh where $25\times 25$
carbon hexagons are commensurate with $23\times 23$ unit cells of Ru \cite%
{martoccia}.

In what follows, we consider a graphene sheet with 2D corrugated profile $%
u_{z}(\boldsymbol{r})$ that forms either due to a lattice misfit with a
substrate or follows its surface corrugation. In order to mix the
unperturbed $\boldsymbol{K}$ and $\boldsymbol{K^{\prime }}$ Bloch states,
such a profile should break the initial translational symmetry in such a way
that both points $\boldsymbol{K}$ and $\boldsymbol{K^{\prime }}$ are
translated into the origin ($\Gamma -$point) within the new ``folded"
Brillouin zone (BZ). Clearly, this can be achieved only if a deformed
graphene sheet that (i) still has a hexagonal Bravais lattice with the angle
$\pi /3$ between two primitive super-periods of equal length, $\boldsymbol{%
\lambda }_{1}$ and $\boldsymbol{\lambda }_{2}$, and (ii) its first
reciprocal lattice vectors with the same length, $\boldsymbol{g}_{1}$ and $%
\boldsymbol{g}_{2}$, are specifically commensurate with the vectors $%
\boldsymbol{K}$ and $\boldsymbol{K^{\prime }}$. As to requirement (i), it is
easy to list all the hexagonal superlattices that are commensurate with the
graphene honeycomb lattice. Indeed, their periods are $\boldsymbol{\lambda }%
_{1}=n\boldsymbol{a}_{1}+m\boldsymbol{a}_{2}$, $\boldsymbol{\lambda }_{2}=-m%
\boldsymbol{a}_{1}+(n+m)\boldsymbol{a}_{2}$, where $n$ and $m$ are arbitrary
integers, $\boldsymbol{a}_{1}$ and $\boldsymbol{a}_{2}$ are the initial
graphene periods. Accordingly, their reciprocal vectors $\boldsymbol{g}$
defined through the corresponding graphene vectors $\boldsymbol{G}$ are: $%
\boldsymbol{g}_{1}=\alpha \boldsymbol{G}_{1}+\beta \boldsymbol{G}_{2}$, $%
\boldsymbol{g}_{2}=\gamma \boldsymbol{G}_{1}+\delta \boldsymbol{G}_{2}$,
where $\alpha =(n+m)/f(n,m),$ $\beta =m/f(n,m),$ $\gamma =-m/f(n,m),$ $%
\delta =n/f(n,m)$; and $f(n,m)=n^{2}+nm+m^{2}$.

From all possible hexagonal superlattices discussed above, not all satisfy
the condition (ii), but only those where $2n+m$ is the multiple of three
\cite{guinea}. Interestingly, in 1D case of $(n,m)$ carbon nanotubes this is
a condition of their gapless (metallic) behavior, while in the present 2D
cases it facilitates opening up the gap, as we discuss below. Such
structures, capable of mapping the graphene $\boldsymbol{K}$ and $%
\boldsymbol{K^{\prime }}$ points to the $\Gamma -$point, can be divided in
three classes: \emph{zigzag, armchair, and chiral}, depending on the
directions of $\boldsymbol{\lambda }$. For the zigzag superlattices, $n=3N, $
$m=0$, so that the periods $\boldsymbol{\lambda }_{1}=3N\boldsymbol{a}{_{1}}$
and $\boldsymbol{\lambda }_{2}=3N\boldsymbol{a}{_{2}}$ are parallel to two
different zigzag directions. This type of superstructures can also be
referred to as $3N\times 3N$, where $N$ is an integer. In the case of
armchair-edged corrugations, $n=m=N$. Here, the super-periods are rotated by
30$^\circ$ relative to the graphene vectors $\boldsymbol{a_1}$ and $%
\boldsymbol{a_2}$. Since the length of the super-periods is larger than that
in graphene by a factor of $\sqrt{3}\,N$, the structures themselves can be
labeled as $\sqrt{3}\,N\times \sqrt{3}\,N$. And, finally, the chiral
superlattices can also be referred to as $\sqrt{f(n,m)}\times \sqrt{f(n,m)}$.

It is worth noticing that the $\sqrt{3}\times \sqrt{3}$ superstructure
breaks the initial translation symmetry in the same way as so-called Kekul%
\'{e} distortion does, where the \emph{in-plane} atomic displacements can be
viewed as a frozen symmetrized combination of transverse optical phonon
modes at $\boldsymbol{K}$-points \cite{lee,gunlycke}. While the Kekul\'{e}
distortion does not break the point group symmetry of the system, D$_{6h}$,
this is not the case for all \emph{out-of plane} 2D corrugations considered
here. Indeed, introducing sine-like corrugations kills at least the
horizontal mirror plane, $\sigma _{h}$, and the resulting point group is
inevitably lower than D$_{6h}$ (actually it can be one of its subgroups: C$%
_{6v}$, D$_{6}$, C$_{6}$, S$_{6}$, D$_{3d}$, C$_{3v}$, D$_{3}$ or C$_{3}$
\cite{milosevic}).

All three classes of periodic ripples, in their simplest form, can be
modeled by the out-of-plane atomic displacements $u_{z}(\boldsymbol{r})$
\begin{equation}
u_{z}(\boldsymbol{r})=\frac{A}{2}\sum_{j=1}^{3}e^{i[\boldsymbol{g}_{j}\cdot (%
\boldsymbol{r}-\boldsymbol{r}_{0})-\varphi ]}+e^{-i[\boldsymbol{g}_{j}\cdot (%
\boldsymbol{r}-\boldsymbol{r}_{0})-\varphi ]},  \label{eq:uz}
\end{equation}%
where $\boldsymbol{r}$ are the undistorted lattice positions, $A$ is the
amplitude, $\boldsymbol{g}_{1},$ $\boldsymbol{g}_{2},$ $\boldsymbol{g}_{3}=-%
\boldsymbol{g}_{1}-\boldsymbol{g}_{2}$ are the first supercell equal-length
reciprocal lattice vectors, each rotated by $2\pi /3$ with respect to one
another, $\boldsymbol{r}_{0}$ is the origin of the coordinate system, and $%
\varphi $ some phase. Below, we will characterize the amplitude $A$ via a
dimensionless ratio $A/\lambda $, where $\lambda =|\boldsymbol{\lambda }%
_{1}|=|\boldsymbol{\lambda }_{2}|=a\sqrt{f(n,m)}$ is the length of a
super-period; $a=|\boldsymbol{a}_{1}|=|\boldsymbol{a}_{2}|$. Note that the
sum of two exponents in Eq.~(\ref{eq:uz}) reduces to the cosine function
when $\varphi =0$, and to the sine function when $\varphi =\pi /2$.

The displacement field $u_{z}\left( \boldsymbol{r}\right) $, Eq.~(\ref{eq:uz}%
), leads to a strain field (quadratic in $u_{z},$ since the linear terms
vanish)
\begin{equation}
\varepsilon _{ij}=\frac{1}{2}\frac{\partial u_{z}}{\partial x_{i}}\times
\frac{\partial u_{z}}{\partial x_{j}},\,\,x_{i}=(x,y),  \label{eq:4}
\end{equation}%
which is also periodic and has non-zero Fourier components of the type $%
\boldsymbol{g}_{i}+\boldsymbol{g}_{j}$. Since such a strain modulates the
interatomic hopping integrals, the electrons of the superlattice should
``feel" strongly not only the components of the potential with the smallest
vectors $\boldsymbol{g}_{i}$, but also those with $\pm \boldsymbol{g}_{i}\pm
\boldsymbol{g}_{i}$. This, in turn, should promote the gap opening due to
enhancing of the second and higher order diffraction effects.

\begin{figure}[h]
\begin{centering}
\includegraphics
[width=9.0cm]{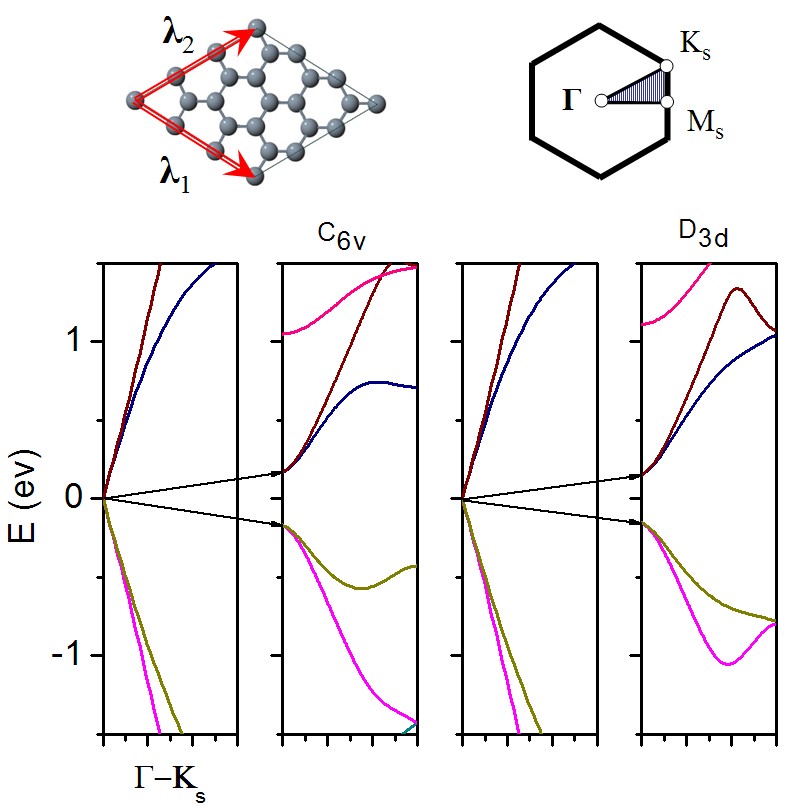}
\caption{(color online) Gap opening  in the zigzag  C$_{6v}$  and  D$_{3d}$ $3\times 3$  superlattices ($\Gamma-K_s$ direction)
 as the ratio  $A/ \lambda$ is increased from 0 to 0.06.  The upper panel shows the periods of
 the superlattices and their irreducible part of the BZ.}
 \label{fig:fig1}
\end{centering}
\end{figure}

The resulting point symmetry of a superstructure imposed by Eq.~(\ref{eq:uz}%
) depends on a choice of the origin $\boldsymbol{r}_{0}$ and phase $\varphi $%
. Table~\ref{tab:table1} lists, for example, the symmetries corresponding to
two choices of $\boldsymbol{r}_{0}$ - either in the center of a hexagon or
on a carbon atom - and two choices of $\varphi $: $\ 0$ and $\pi /2$.
Although all the superstructures listed in Table~\ref{tab:table1} have
identical in shape (hexagonal) BZs, the irreducible part of the latter can
differ in going from one structure to another. Thus, the irreducible part
for the C$_{6v}$ and D$_{3d}$ superlattices is geometrically identical to
that for graphene, whereas for C$_{6}$, S$_{6}$, C$_{3v}$ and D$_{3}$ is
twice and for C$_{3}$ four time larger than that for graphene \cite%
{milosevic}. For the C$_{6v}$ and D$_{3d}$ structures, we denote the
high-symmetry points at the corners of the superlattice BZ as K$_{s}$, and
at centers of the edges as M$_{s}$, Fig.~\ref{fig:fig1}.

\begin{table}[tbp]
\caption{Point groups of 2D corrugated structures as a function of $%
\boldsymbol{r}_{0}$ and $\protect\varphi $ chosen in Eq.~(\protect\ref{eq:uz}%
). Note that the displacements $u_{z}$'s (\protect\ref{eq:uz}) are
identically zero for the $N=\protect\sqrt{3}\times \protect\sqrt{3}$
structure when $\boldsymbol{r}_{0}$ is in the center of a hexagon and $%
\protect\varphi =0,\protect\pi /2$, so that the structures of this family
appear only for $N>1$.}
\label{tab:table1}
.
\begin{ruledtabular}
     \begin{tabular}{|c|c|c|}
  $zigzag \,\,3N \times 3N$ &center of hexagon&  carbon atom \\
  \tableline
   $\varphi=0$&C$_{6v}$ &C$_{3v}$\\
   $\varphi=\pi/2 $&D$_{3d}$&C$_{3v}$\\
 \tableline
$armchair  \,\,\sqrt{3}\, N \times \sqrt{3}\, N$ & &\\
 \tableline
  $\varphi=0$&C$_{6v}$ ($N>$1)&C$_{3v}$\\
  $\varphi=\pi$/2 &D$_{3d}$($N>$1) &D$_{3}$\\
  \tableline
$chiral \,\,\sqrt{f}\times \sqrt{f}
$ & &\\
 \tableline
  $\varphi=0$&C$_{6}$ &C$_{3}$\\
  $\varphi=\pi$/2 &S$_{6}$ &C$_{3}$\\

     \end{tabular}
   \end{ruledtabular}
\end{table}
To illustrate the suggested concept, we have performed band structure
calculations for the corrugated graphene sheets using density functional
theory implemented in the ABINIT package \cite{abinit} (for computational
details, see Supplemental Material \ref{sec:suppl}.) Among the groups listed
in Table~\ref{tab:table1}, only C$_{6v}$, D$_{3d}$, C$_{6}$ and S$_{6}$
contain the symmetry elements that interchange the A and B sublattices. For
such groups, the gap opens up solely due to inter-valley mixing: the
distortion breaks the 4-fold degeneracy at $\Gamma $ point of a flat
graphene into two 2-fold degeneracies. In the case of armchair C$_{6v}$ and D%
$_{3d}$ $\ 3\times 3$ superlattices, the gap reaches substantial values of
0.33 and 0.30~eV, respectively, as the $A/\lambda $ increases from $0$ to $%
0.06$ (Fig.~\ref{fig:fig1}). With the constraint $A/\lambda =0.06$, these
values are reduced by a factor of three in passing to the armchair $2\sqrt{3}%
\times 2\sqrt{3}$ superlattices having the same symmetries. They are further
reduced by a factor $2-3$ in going to the chiral C$_{6}$ and S$_{6}$ $\sqrt{%
21}\times \sqrt{21}$ structures corresponding to $n=1$ and $m=4$. Clearly,
this progressive narrowing of the gap is due to increasing period $\lambda $%
, because the ratio $A/\lambda $ remains the same.

\begin{figure}[h]
\begin{centering}
\includegraphics
[width=9cm]{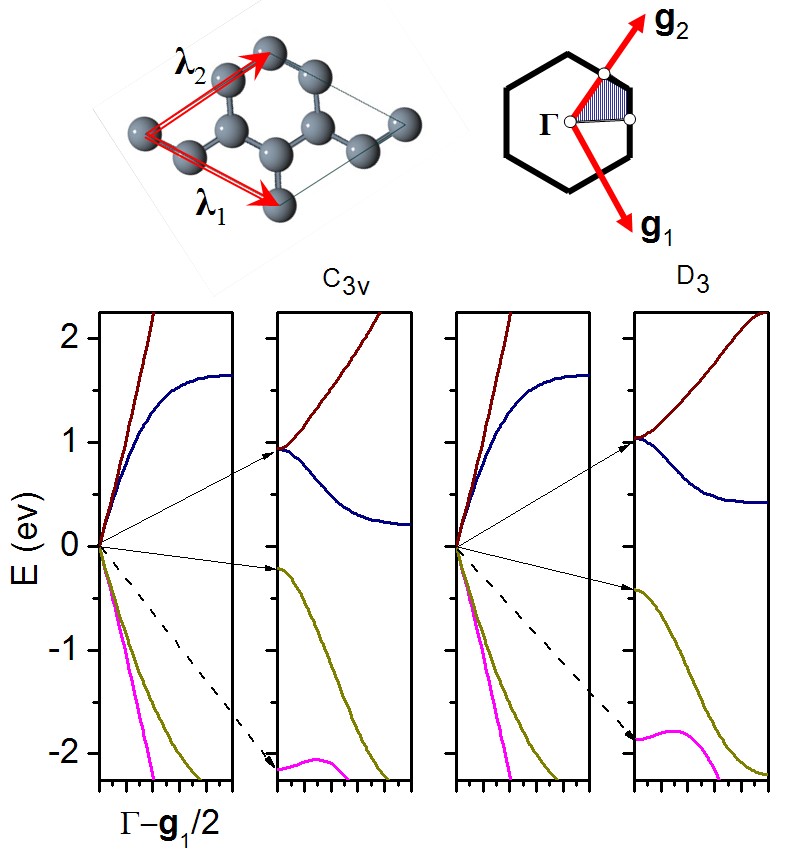}
\caption{(color online) Same as in Fig. 1,  but  for the armchair   C$_{3v}$  and  D$_{3}$  $\sqrt{3} \times \sqrt{3}$  superlattices
in the $\Gamma-\boldsymbol{g}_{1}/2$ direction.
}\label{fig:fig2}
\end{centering}
\end{figure}
\begin{figure}[h]
\begin{centering}
\includegraphics
[width=9cm]{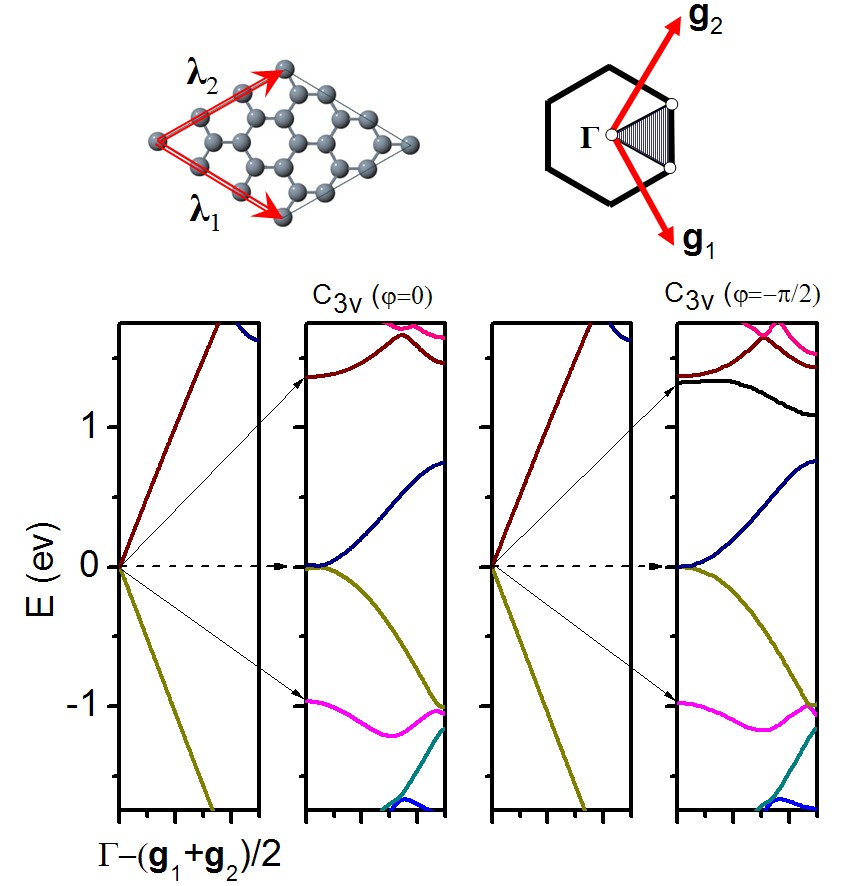}
\caption{(color online)  Same as in Fig.1, but for  the
C$_{3v}$  superlattices in the $\Gamma-(\boldsymbol{g}_{1}+ \boldsymbol{g}_{2})/2$ direction.
 }\label{fig:fig3}
\end{centering}
\end{figure}
\begin{figure}[h]
\begin{centering}
\includegraphics
[width=9cm]{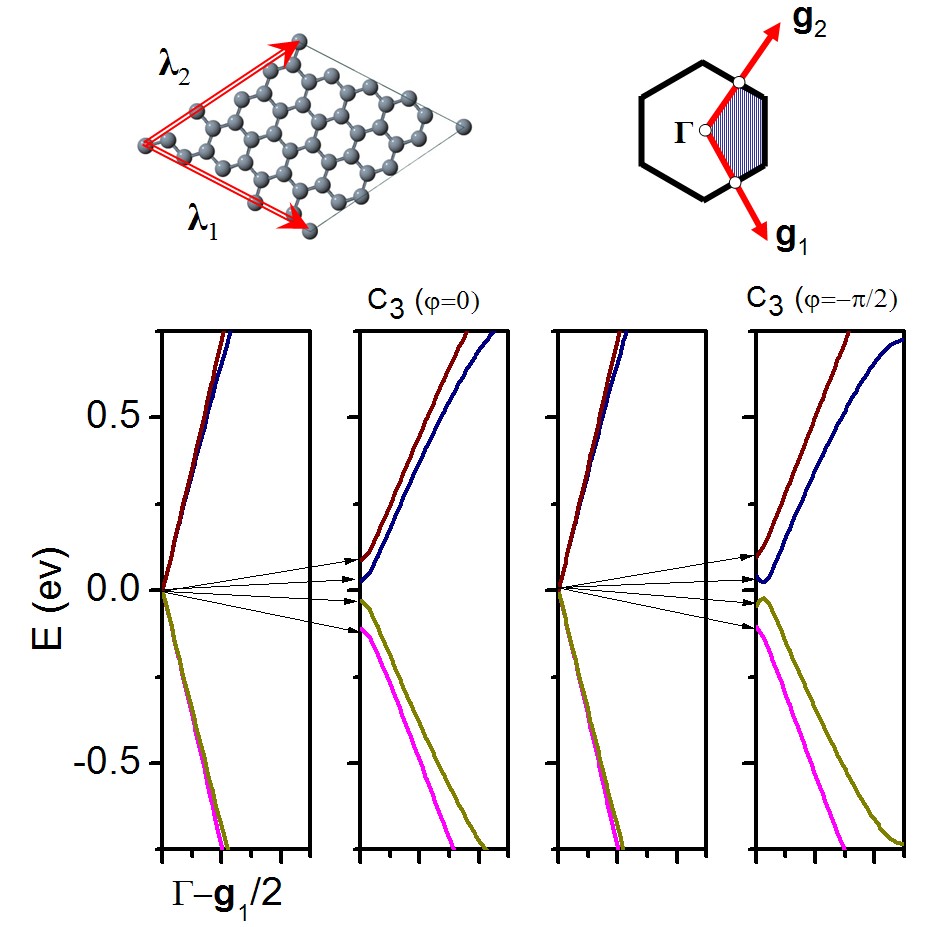}
\caption{(color online)  Same as in Fig.1, but for the chiral  C$_{3}$ $\sqrt{21} \times \sqrt{21}$   superlattices
in the  $\Gamma-\boldsymbol{g}_{1}/2$ direction.
}\label{fig:fig4}
\end{centering}
\end{figure}
The corrugations associated with the symmetries C$_{3v}$, D$_{3}$ and C$_{3}$
(Table~\ref{tab:table1}) break not only the translational but also the
sublattice symmetry. In such an event, both the inter-valley and
intra-valley mixing effects contribute to the reconstruction of the band
structure in the vicinity of the Fermi level. In the case of armchair C$%
_{3v} $ and D$_{3}$ $\sqrt{3}\times \sqrt{3}$ structures, for example, the
four-fold degenerate state at $\Gamma $ splits into two single levels and a
two-fold degenerate state, Fig.~\ref{fig:fig2}. The 2-fold degenerate state
is above the Fermi level, so that the gap opens up and becomes indirect when
the $A/\lambda $ exceeds some critical value (as in Fig.~\ref{fig:fig2}).
Although in the zigzag C$_{3v}$ $3\times 3$ superlattices the 4-fold
degenerate state also splits into three states, here the two-fold degenerate
state lies between the two single levels and the gap does not open up at
all, Fig.~\ref{fig:fig3}. This result is quite unexpected, it shows that
breaking the inversion symmetry is not a \emph{sufficient} condition for the
gap opening. Note that the principal possibility for graphene lattice to
have zero energy gap in the absence of inversion symmetry was discussed
earlier in Ref.~\cite{kishigi}.

It is interesting that in going over from $\sqrt{3}\times \sqrt{3}$ to $%
\left( 3\sqrt{3}\right) \times \left( 3\sqrt{3}\right) $ structures with the
same point symmetries (C$_{3v}$ and D$_{3}$), the character of splitting of
the 4-fold degenerate state remains but only for very small corrugation
parameters $A/\lambda \leq 0.005$. As the $A/\lambda $ is further increased,
the 2-fold degeneracy is lifted and the structures exhibits four single
levels. We should stress that if only the lowest order effects of the
corrugations were considered (quadratic in $A/\lambda $), the splitting of
the four-degenerate level would follow the ratios obtained from symmetry
considerations. In the present calculations, however, the effects of
corrugation are included up to an infinite order, so that the character of
splitting is not in one-to-one correspondence with the symmetry of deformed
structures. The 4-fold degeneracy is completely lifted at any $A/\lambda $
for the chiral superlattices with C$_3$ symmetries, as exemplified by the C$%
_{3}$ $\sqrt{21}\times \sqrt{21}$ structure in Fig.~\ref{fig:fig4}. Here, a
modest gap on the order of 0.1 eV opens up as $A/\lambda $ reaches $0.06$.
Note that once the gap is opened at $\Gamma$, it does not necessarily
remains the direct gap, the conduction band minimum may move around in the
BZ, and the gap may become indirect, as it is seen from the Fig.~\ref%
{fig:fig4} for the case $\varphi =\pi /2$. The above results provide
exhaustive classification of possible scenarios for gap opening in graphene
subject to the two-dimensional corrugation. It would be interesting to check
the above predictions in experiments with various substrates producing
planar corrugations of a top graphene layer.

\textbf{Supplement: Calculation Method}. The DFT calculations for both flat and
corrugated graphene have been performed using ABINIT \cite{abinit} within
the local density (LDA) approximation. Approximately the same k-point
density for Monkhorst-Pack $\boldsymbol{k}$-point grid \cite{kpts} was used
for all superstructures; roughly it corresponds to a 16$\times $16$\times $1
grid for a flat graphene with 2 atoms per unit cell. The sheets have been
simulated by a slab-supercell approach with the inter-planar distances of $%
30a_{B}$ to ensure negligible wave function overlap between the replica
sheets. The plane waves pseudopotentials have been chosen in the form of
Troullier-Martins \cite{trou91} where carbon 2s and 2p electrons have been
considered as valence states. For the plane-wave expansion of the valence
and conduction band wave-functions, a cutoff energy was chosen to be 80 Ry.


\begin{thebibliography}{99}
\bibitem{bolotin} K. I. Bolotin \emph{et al}., Solid State Comm. \textbf{146}%
, 351 (2008).

\bibitem{morozov} S. V. Morozov \emph{et al}., Phys. Rev. Lett. \textbf{100}%
, 016602 (2008).

\bibitem{dusari2011} S. Dusari, J. Barzola-Quiquia, P. Esquinazi, and
N.~Garcia, Phys. Rev. B \textbf{83}, 125402 (2011).

\bibitem{leeThrmlG} J.-U. Lee, D. Yoon, H. Kim, S.W. Lee, and H. Cheong,
Phys. Rev. B \textbf{83}, 081419(R) (2011).

\bibitem{berger} C. Berger \emph{et al}., Science \textbf{312}, 1191 (2006).

\bibitem{han} M. Y. Han, B. Ozyilmaz, Y. Zhang, P. Kim, Phys. Rev. Lett.
\textbf{98}, 206805 (2007).

\bibitem{ponomarenko} L.A. Ponomarenko \emph{et al}., Science \textbf{320},
356 (2008).

\bibitem{zhou} S. Y. Zhou \emph{et al.}, Nature Mat. \textbf{6}, 770 (2007).

\bibitem{gioavannetti} G. Giovannetti, P.A. Khomyakov, G. Brocks, P.J.
Kelly, J. van den Brink, Phys. Rev. B \textbf{76}, 073103 (2007).

\bibitem{yu} D. Yu \emph{et al}., Nano Res. \textbf{1}, 56 (2008).

\bibitem{vanevic} M. Vanevi\'{c}, V. M. Stojanovi\'{c}, and M. Kindermann,
Phys. Rev. B \textbf{80}, 045410 (2009).

\bibitem{liu} W. Liu, Z.F. Wang, Q. W. Shi, J. Yang, and F. Liu, Phys. Rev.
B \textbf{80}, 233405 (2009)

\bibitem{baskin} A. Baskin and P. Kr\'{a}l, Nature Sci. Reports \textbf{1},
36 (2011).

\bibitem{chernozatonskii} L. A. Chernozatonskii \emph{et al.}, JETP Letters
\textbf{85}, 77 (2007).

\bibitem{ito} J. Ito and A. Natori, J. Appl. Phys. \textbf{103}, 113712
(2008).

\bibitem{balog} R. Balog \emph{et al.}, Nature Mat. \textbf{9}, 315 (2010).

\bibitem{xu} Z. Xu and K. Xue, Nanotechnology \textbf{21}, 045704 (2010).

\bibitem{snyman} I. Snyman, Phys. Rev. B \textbf{80}, 054303 (2009).

\bibitem{savel'ev} S. E. Savel'ev and A.S. Alexandrov, Phys. Rev. B \textbf{%
84}, 035428 (2011).

\bibitem{lee} S.-H. Lee \emph{et al}. ACS Nano \textbf{5}, 2964 (2011).

\bibitem{naubra1d} I.I. Naumov and A.M. Bratkovsky, arXiv:1104.0314v1
[cond-mat.mes-hall].

\bibitem{pacoNP10} F. Guinea, M. I. Katsnelson, and A. K. Geim, Nature Phys.
\textbf{6}, 30 (2010).

\bibitem{emtsev} K. V. Emtsev, F. Speck, Th. Seyller, L. Ley, and J.D.
Riley, Phys. Rev. B \textbf{77}, 155303 (2008).

\bibitem{kim1} S. Kim, J. Ihm, H. J. Choi, and Y.-W. Son, Phys. Rev. Lett.
\textbf{100}, 176802 (2008).

\bibitem{riedl} C. Riedl, C. Coletti and U. Starke, J. Phys. D \textbf{43},
374009 (2010).

\bibitem{qi} Y. Qi, S. H. Rhim, G. F. Sun, M. Weinert, and L. Li, Phys. Rev.
Lett. \textbf{105}, 085502 (2010).

\bibitem{preobrajenski} A. B. Preobrajenski, M. L. Ng, A. S. Vinogradov, and
N. M{\aa }rtensson, Phys. Rev. B \textbf{78}, 073401 (2008).

\bibitem{martoccia} D. Martoccia, \emph{et al.}, Phys. Rev. Lett. \textbf{101%
}, 126102 (2008).

\bibitem{ma} H.-F. Ma, M. Thomann, J. Schmidlin, S. Roth, M. Morscher, and
T. Greber, Front. Phys. China \textbf{5}, 387 (2010).

\bibitem{guinea} F. Guinea and T. Low, Phil. Trans. R. Soc. A \textbf{368}
5391 (2010)

\bibitem{gunlycke} D. Gunlycke, H.M. Lawler and C. T. White, Phys. Rev. B
\textbf{75}, 085418 (2007).

\bibitem{milosevic} I. Milo\u{s}evi\'{c}, B. Nikoli\'{c}, M. Damnjanovi\'{c}
and M. Kr\u{c}marz, J. Phys. A \textbf{31},3625 (1998).

\bibitem{kishigi} K. Kishigi, R. Takeda, and Y. Hasegawa, J. Phys. Conf.
Ser. \textbf{132}, 101205 (2005).

\bibitem{abinit} X. Gonze \emph{et al}., Comp. Mater. Sci. \textbf{25}, 478
(2002).

\bibitem{kpts} H. J. Monkhorst and J. D. Pack, Phys. Rev. B \textbf{13},
5188 (1976).

\bibitem{trou91} N. Troullier and J. L. Martins, Phys. Rev. B \textbf{43},
1993 (1991).
\end{thebibliography}
\end{document}